\documentclass[12pt,preprint]{aastex62}
\usepackage{graphicx}
\usepackage{longtable}
\usepackage{multirow}
\usepackage{url}
\usepackage{subfigure}
\usepackage{amsmath}
\usepackage{lipsum}

\begin{document}
\title{A New Method to Measure Hubble Parameter $H(z)$ using Fast Radio Bursts}
\author[0000-0002-7555-0790]{Q. Wu}
\affiliation{School of Astronomy and Space Science, Nanjing
University, Nanjing 210093, China}

\author[0000-0002-7555-0790]{H. Yu}
\affiliation{Department of Astronomy, School of Physics and Astronomy, Shanghai Jiao Tong University, Shanghai, China}

\author[0000-0002-7555-0790]{F. Y. Wang}
\affiliation{School of Astronomy and Space Science, Nanjing
University, Nanjing 210093, China} \affiliation{Key Laboratory of
Modern Astronomy and Astrophysics (Nanjing University), Ministry of
Education, Nanjing 210093, China} \email{{*}fayinwang@nju.edu.cn}

\begin{abstract}
The Hubble parameter $H(z)$ is directly related to the expansion of
our Universe. It can be used to study dark energy and constrain
cosmology models. In this paper, we propose that $H(z)$ can be
measured using fast radio bursts (FRBs) with redshift measurements.
We use dispersion measures contributed by the intergalactic medium,
which is related to $H(z)$, to measure Hubble parameter. We find
that 500 mocked FRBs with dispersion measures and redshift
information can accurately measure Hubble parameters using Monte
Carlo simulation. The maximum deviation of $H(z)$ from standard
$\Lambda$CDM model is about 6\% at redshift $z= 2.4$. We
also test our method using Monte Carlo simulation.
Kolmogorov-Smirnov (K-S) test is used to check the simulation. The
$p$-value of K-S test is 0.23, which confirms internal consistency
of the simulation. In future, more localizations of FRBs make it as
an attractive cosmological probe.
\end{abstract}


\section{Introduction}
Since the creative work of Edwin Powell Hubble in 1929
\citep{Hubble1929}, the fact that our universe is evolving and under
expansion is established. Seven decades later, the accelerating
expansion of our universe was discovered by measurements of type Ia
supernovae (SNe Ia) \citep{Riess1998,Perlmutter1999}, which changed
our understanding of the universe again. The new findings have
encouraged people to investigate the mysterious component, which is
called dark energy, via several different ways.

The cosmic expansion rate, expressed in terms of Hubble parameter
$H(z)=\dot{a}(t)/a(t)$ with scale factor $a(t)$, is a powerful
cosmological probe \citep[for reviews, see][]{Zhang2010}. In flat
$\Lambda $CDM cosmology, $H(z)$ can be expressed as
\begin{equation}
H(z) = H_{0}\sqrt{\Omega_{\Lambda}+\Omega_{m}(1+z)^3},
\end{equation}
where $H_{0}$ is the Hubble constant, $\Omega_{\Lambda}$ is the
vacuum energy density fraction, and $\Omega_{m}$ is the matter
density fraction.

The cosmic expansion rate $H(z)$ is a powerful tool to study
cosmological parameters \citep{Samushia2006,Farooq2017,Tu2019},
cosmological deceleration to acceleration transition
\citep{Farooq2013,Yu2018,Jesus2018}, the Hubble constant
\citep{Busti2014,Chen2017,Wang2017,Yu2018} and cosmic curvature
\citep{Clarkson2007,Clarkson2008,Yu2016}. Several methods have been
proposed to measure $H(z)$. The first one is the differential age
method, which was first put forward by \cite{Jimenez2002}. Some
efforts have been performed
\citep{Stern2010,Moresco2012,Liu2012,Zhang2014,Ratsimbazafy2017}.
However, it is difficult to select galaxies that can act as ``cosmic
chronometers'' and determine the age of a galaxy. This method is
relying on population synthesis simulations based on standard
physics and cosmology. The second one is radial baryon acoustic
oscillation (BAO) size method
\citep{Blake2012,Font-Ribera2014,Delubac2015,Alam2017}. However, the
Hubble parameter degenerates with the comoving distance and the
derived $H(z)$ depends on the assuming cosmological model in this
method. More recently, \cite{Amendola2019} proposed that $H(z)$ can
be derived by measuring the power spectrum of density contrast and
peculiar velocities of supernovae. In this paper, we propose that
$H(z)$ can be measured using fast radio bursts (FRBs) with redshift
measurements.

FRBs are very bright and short bursts with high brightness
temperatures \citep{Lorimer2007,Thornton2013,Champion2016,Katz2018,
Platts2019,Cordes2019}, which are considered to have cosmological
origin. One of the significant characteristics is that FRBs have
large dispersion measure (DM), which is proportional to the integral
of free electron density along the line of sight from the source to
the observer. DM can be used as a cosmological probe. At present,
more than 100 FRBs have been detected. Twenty of them are repeaters,
most of which were discovered by the Canadian Hydrogen Intensity
Mapping Experiment (CHIME) \citep{The CHIME/FRB
Collaboration2019a,The CHIME/FRB Collaboration2019b, Fonseca2020}.
So far, five FRBs have been localized and two of them are repeating
bursts. A direct localization of FRB 121102 by Very Large Array
(VLA) at redshift $z=0.19$ \citep{Chatterjee2017,Tendulkar2017}
confirmed the cosmological origin of this source. Recently, the
repeating FRB 180916.J0158+65 is localized to a nearby spiral galaxy
($z=0.0337$) \citep{Marcote2020}. For one-off FRBs, FRB 180924 was
located in a position 4 kpc from the center of an early-type spiral
galaxy at a redshift of 0.32 \citep{Bannister2019}. FRB 190523 was
located in a few-arcsecond region containing a massive galaxy at
redshift 0.66 \citep{Ravi2019}. FRB 181112 was localized in a galaxy
at redshift 0.47 \citep{Prochaska2019}. More and more FRBs with
measured redshifts will be detected in future based on the high rate
of FRBs, which reaches $\rm~10^4 \ sky^{-1} day^{-1}$
\citep{Thornton2013}. Meanwhile, the CHIME telescope with an
effective field of view about 250 square degrees can detect FRBs
with an unexpected rate. It would provide a large data sample for
measuring Hubble parameter. FRBs with redshift and DM measurements
will be useful cosmological and astrophysical probes, including
measuring baryon number density
\citep{Zheng2014,Deng2014,Walters2019,Wei2019}, measuring cosmic
proper distance \citep{Yu2017} and the cosmological parameters
\citep{Zhou2014,Gao2014,Walters2018,Li2018,Jaroszynski2019}, probing
compact dark matter \citep{Munoz2016,Wang2018} and testing
Einstein's weak equivalence principle (WEP)
\citep{Wei2015,Yu2018a,Xing2019}. \cite{Kumar2019} gave a quantitive
estimation for the systematics control needed for using FRB
dispersion measures as distance probe.

In this paper, we propose a new method to measure Hubble parameter
using FRBs with redshift and DM measurements. This paper is
organized as follows. In section \ref{sec2}, we give an introduction
of the method. In section \ref{sec3}, our method is tested using
simulated FRB sample. Conclusions and discussion will be given in
section \ref{sec4}.

\section{Method for measuring $H(z)$}\label{sec2}
The observed $\rm DM_{obs}$ includes contributions from the
intergalactic medium (IGM), the Milky Way, and local environment
(including the host galaxy and the source). It can be expressed as
\begin{equation}\label{DM_obs}
\rm DM_{obs} =DM_{IGM}+DM_{MW}+\frac{DM_{loc}}{1+\emph{z}}.
\end{equation}
$\rm DM_{IGM}$ is the only parameter which contains the information
of Hubble parameter in this equation. The mean dispersion
measure caused by the inhomogeneous IGM is given by
\begin{equation}\label{DM_igm}
{\rm \langle DM_{IGM} \rangle } = {\rm A\Omega_b} H_0\int^z_0\frac{F(z)}{E(z^\prime)}dz^\prime,
\end{equation}
where $E(z)=H(z)/H_{0}$, $F(z) = (1+z)f_{\rm IGM}(z)f_{\rm e}(z)$,
and $A=3c/8\pi Gm_{\rm p}$. $\Omega_{\rm b}$ is the cosmic baryon
mass density fraction, $m_p$ is the mass of proton, and $f_{\rm
IGM}$ is the fraction of baryon mass in the IGM. $f_{\rm e} = Y_{\rm
H}X_{\rm e,H}(z)+\frac{1}{2}Y_{\rm He}X_{\rm e,He}(z)$, $Y_{\rm
H}=3/4$ and $Y_{\rm He}=1/4$ are the mass fractions of hydrogen and
helium, respectively. $X_{\rm e,H}$ and $X_{\rm e,He}$ are the
ionization fractions of intergalactic hydrogen and helium,
respectively. At $z<3$, hydrogen and helium are fully ionized. So
$X_{\rm e,H}=X_{\rm e,He}=1$. The values of $f_{\rm IGM}$ are $0.82$
and $0.9$ at $z<0.4$ and $z>1.5$, respectively \citep{Shull2012}.
In the redshift range $0.4<z<1.5$, $f_{\rm IGM}$ may
change linearly at $0.4<z<1.5$ with a random deviation of 0.04
\citep{Zhou2014}. 


We assume that the data is divided into several redshift bins and the
averaged redshifts $\langle z \rangle$ and dispersion measures $\langle \rm DM_{IGM} \rangle$
are known for each of them. Differentiating equation (\ref{DM_igm}), Hubble parameter
can be expressed as
\begin{equation}\label{eq_Hz}
H(z)={\rm A\Omega_b}H_0^2F(z)\frac{\Delta z}{\Delta \rm DM_{IGM}},
\end{equation}
where ${\rm \Delta} z$ and $\rm \Delta DM_{IGM}$ are the differences
of $z$ and DM$_{\rm IGM}$ between two adjacent bins, respectively.
The reciprocal of $\Delta z / \rm \Delta DM_{IGM}$ represents the
first derivative of $\rm DM_{IGM}$ with respect to redshift
$z$. Only $\rm DM_{IGM}$ depends on the Hubble parameter in
equation (\ref{DM_obs}). Other components must be subtracted from
$\rm DM_{obs}$. For $\rm DM_{loc}$, which contains the $\rm
DM_{host}$ and $\rm DM_{source}$, one method has been proposed to
determine it using low-$z$ FRBs \citep{Yang2016}. \cite{Zhang2019}
studied the $\rm DM_{host}$ using IllustrisTNG simulation and found
the value of $\rm DM_{host}$ is almost independent of redshift for
non-repeating FRBs at $z<1.5$. $\rm DM_{source}$ depends on the
progenitors of FRBs. If FRBs are born in binary neutron star systems
\citep{Wang2016,Zhang2020,Wang2020}, the value of $\rm DM_{source}$
is small. $\rm DM_{loc}$ is divided by a $(1+z)$ factor due
to the time delay. Here $\rm DM_{IGM}$ increases with redshift, so
$\rm DM_{loc}$ is not important at high redshifts. $\rm \langle
DM_{loc}\rangle = 200~{pc/cm^3}$ is assumed \citep{Yu2017}. We
subtract $\rm DM_{loc}$ and leave its uncertainty into the total
uncertainty $\sigma_{\rm total}$. The total uncertainty $\sigma_{\rm
total}$ is
\begin{equation}\label{sigma_tot}
\sigma_{\rm total}^2 = \sigma_{\rm obs}^2 + \sigma_{\rm MW}^2 +\sigma_{\rm IGM}^2 + \left(\frac{\sigma_{\rm loc}}{1+z}\right)^2  .
\end{equation}
Since the measurements of DM are accurate, the uncertainties of $\rm
DM_{obs}$ and $\rm DM_{WM}$ can be omitted compared with the much
larger uncertainties of $\rm DM_{ loc}$ and $\rm DM_{IGM}$.
Following \cite{Thornton2013} and numerical simulations of
\cite{McQuinn2014}, we choose $\rm \sigma_{DM_{loc}}=100 ~pc/cm^3$
in analysis. Due to the inhomogeneity of IGM, the uncertainty of $\rm
DM_{IGM}$ is related to redshift, which can be expressed as
\citep{Kumar2019}
\begin{equation}\label{eq_sigma_DM_IGM}
\frac{\sigma_{\rm DM_{IGM}}}{\rm DM_{IGM}} = \frac{20\%}{\sqrt{z}}.
\end{equation}
Fortunately, if there are several FRBs from different sightlines but
in a narrow redshift bin, uncertainty of averaged $\rm
\langle DM_{IGM} \rangle $ decreases by the square root of the
number of FRBs in this bin.

The $\rm DM_{MW}$ can be subtracted from pulsar observations
\citep{Taylor1993,Manchester2005}. Some models have been proposed to
describe the distribution $\rm DM_{WM}$ \citep{Cordes2002,Yao2017}.
Therefore, the value $\rm DM_E=DM_{obs}-DM_{MW}$ is
\begin{equation}\label{eq_DM_E}
{\rm DM_E }= {\rm DM_{IGM}}+\frac{\rm DM_{loc}}{1+z} = {\rm
A\Omega_b} H_0\int^z_0\frac{F(z)}{E(z^\prime)}dz^\prime + \frac{\rm
DM_{loc}}{1+z}.
\end{equation}

According the results of \cite{Zhang2019}, we assume that the value
of $\rm DM_{loc}$ does not evolve with redshift significantly. The
uncertainty of $\rm DM_{E}$ is
\begin{equation}\label{eq_dDM_E1}
\Delta {\rm DM_E} = \Delta {\rm DM_{IGM}}-\frac{{\rm DM_{loc}}\Delta z}{(1+z)^2} .
\end{equation}
Thus, we can calculate $\frac{\Delta z}{\Delta {\rm DM_{IGM}}}$ from
\begin{equation}\label{eq_dDM_E2}
\frac{\Delta z}{\Delta {\rm DM_{IGM}}} = \frac{1}{\frac{\Delta
z}{\Delta {\rm DM_E}} + \frac{{\rm DM_{loc}}}{(1+z)^2}}.
\end{equation}
The effect of $\rm DM_{loc}$ is not important at high
redshifts. From equation (\ref{DM_igm}), the uncertainty of $\rm
DM_{IGM}$ is
\begin{equation}\label{eq_dDM_IGM}
\Delta {\rm DM_{IGM}} = \Delta  {\rm DM_E} + \frac{{\rm DM_{loc}}\Delta
    z}{(1+z)^2} = \frac{{\rm A\Omega_b}H_0 F(z)\Delta z}{E(z)}.
\end{equation}
Then we can derive the error of $H(z)$ as
\begin{equation}\label{eq_sigma}
\left(\frac{\sigma_{H(z)}}{H(z)}\right)^2 = \left(\frac{\sigma_{\Omega_bH_0^2}}{\Omega_bH_0^2}\right)^2 + \left(\frac{\sigma_{F(z)}}{F(z)}\right)^2 + \left(\frac{\sigma_{\frac{\Delta z}{\Delta {\rm DM_{IGM}}}}}{\frac{\Delta z}{\Delta {\rm DM_{IGM}}}}\right)^2.
\end{equation}
Here we assume that the uncertainty of $F(z)$ is 0.04,
of $f_{\rm IGM}$ is 0.04  \citep{Zhou2014}, and of $\sigma_{\Omega_bH_0^2}$ is 0.01.

Assuming that a FRB data set $(z, \rm DM_{obs})$ is available, 
we use the following steps to derive the Hubble parameter $H(z)$.
(i) Deriving the data set $(z,\rm DM_{E})$ by subtracting $\rm DM_{MW}$;
(ii) Separating the data set $(z,\rm DM_{E})$ into five redshift bins, and then
calculating the averaged redshifts, the averaged $\rm DM_{E}$ and
the uncertainty of $\rm DM_{E}$. Then we obtain a data set $\langle
z\rangle$, $\langle \rm DM_{E}\rangle$ and $\rm\sigma_{DM_{E}}$;
(iii) Deriving $\frac{\Delta z}{\Delta {\rm DM_{E}}}$, then
$\frac{\Delta z}{\Delta {\rm DM_{IGM}}}$ can be calculated using
equation (\ref{eq_dDM_E2}). Equation (\ref{eq_Hz}) can be used to
calculate the Hubble parameters $H(z)$. The uncertainty of $H(z)$ is
derived from equation (\ref{eq_sigma}).

\section{Monte Carlo Simulations and Results}\label{sec3}

\begin{figure}[htbp]
    \centering

    \subfigure[]{
        \begin{minipage}[t]{0.48\linewidth}
            \centering
            \includegraphics[width=3.8in]{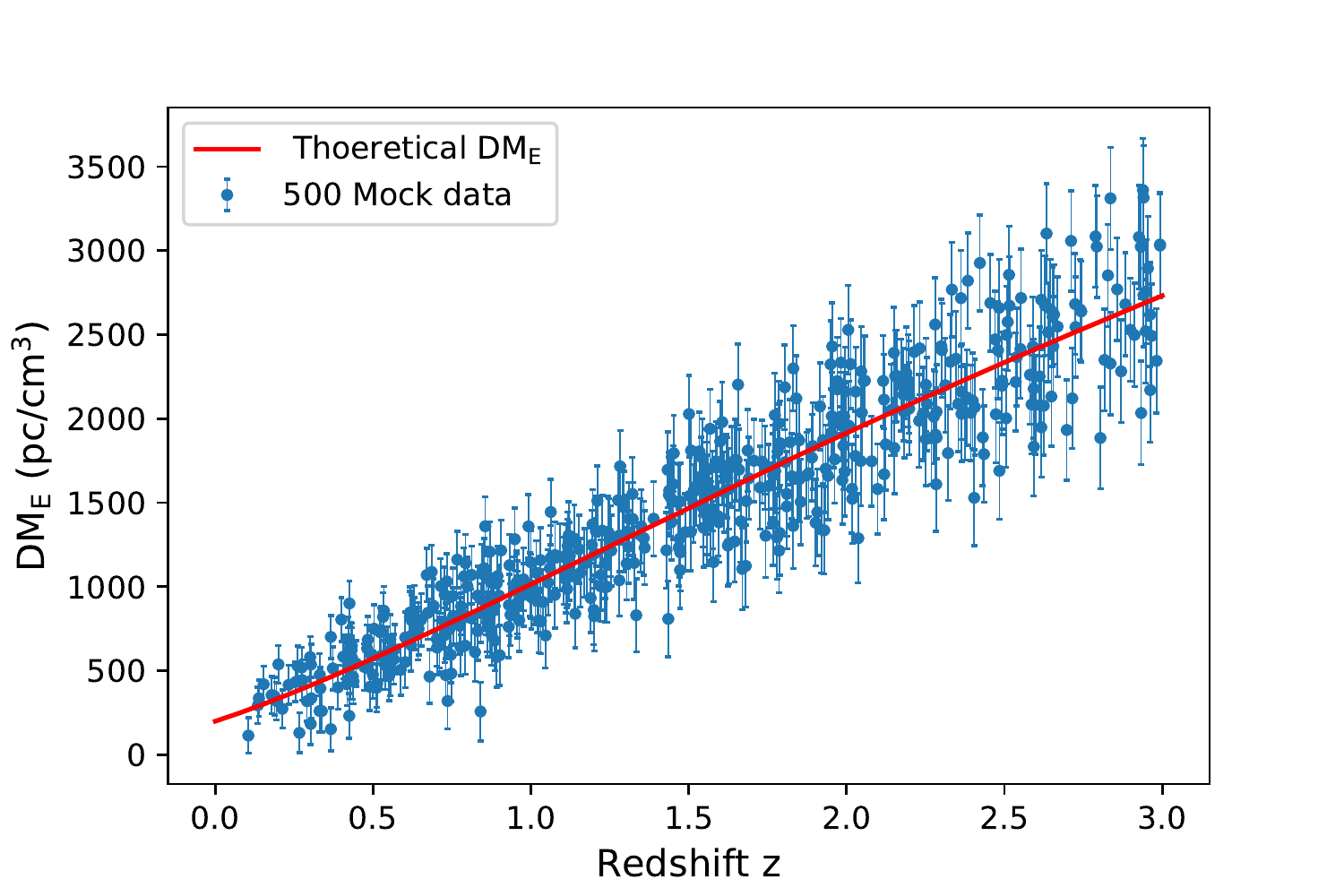}
        \end{minipage}%
    }
    \subfigure[]{
        \begin{minipage}[t]{0.48\linewidth}
            \centering
            \includegraphics[width=3.8in]{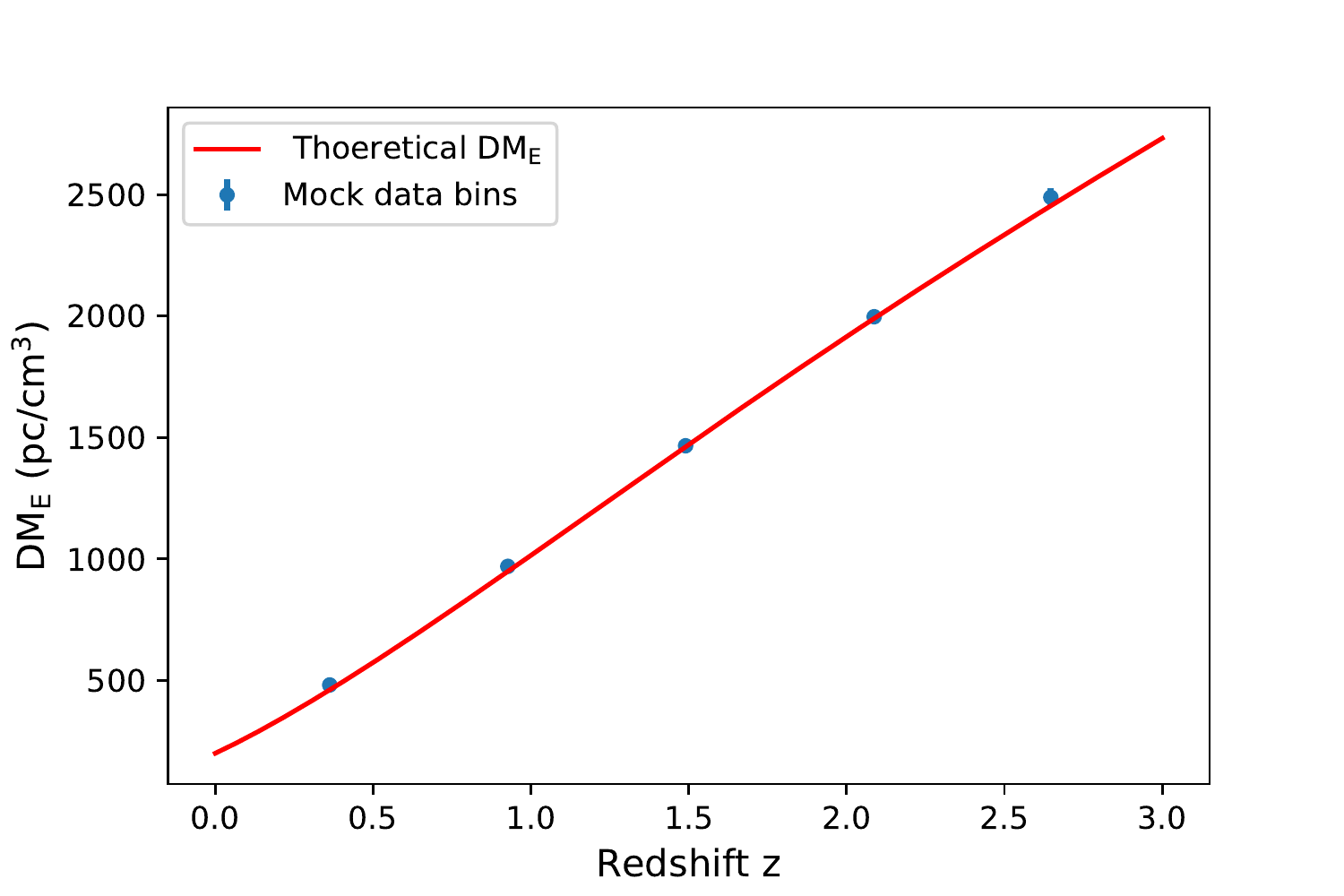}
        \end{minipage}%
    }
    \centering
    \caption{The left panel shows the 500 mock DMs of FRBs (blue dots) with $1\sigma$ errors
        and the theoretical $\rm DM_{E}(z)$ function (red line). The
        right panel shows the five bins of the average redshift
        $\langle z\rangle$ and the corresponding $\rm DM_{E}$.
        The error of $\rm DM_{E}$ is about
        20~$\rm pc/cm^3$. The red line is the theoretical DM$_{\rm E}(z)$
        function.}
    \label{fig1}
\end{figure}

\begin{figure}
    \centering
    \includegraphics[width=0.6\textwidth]{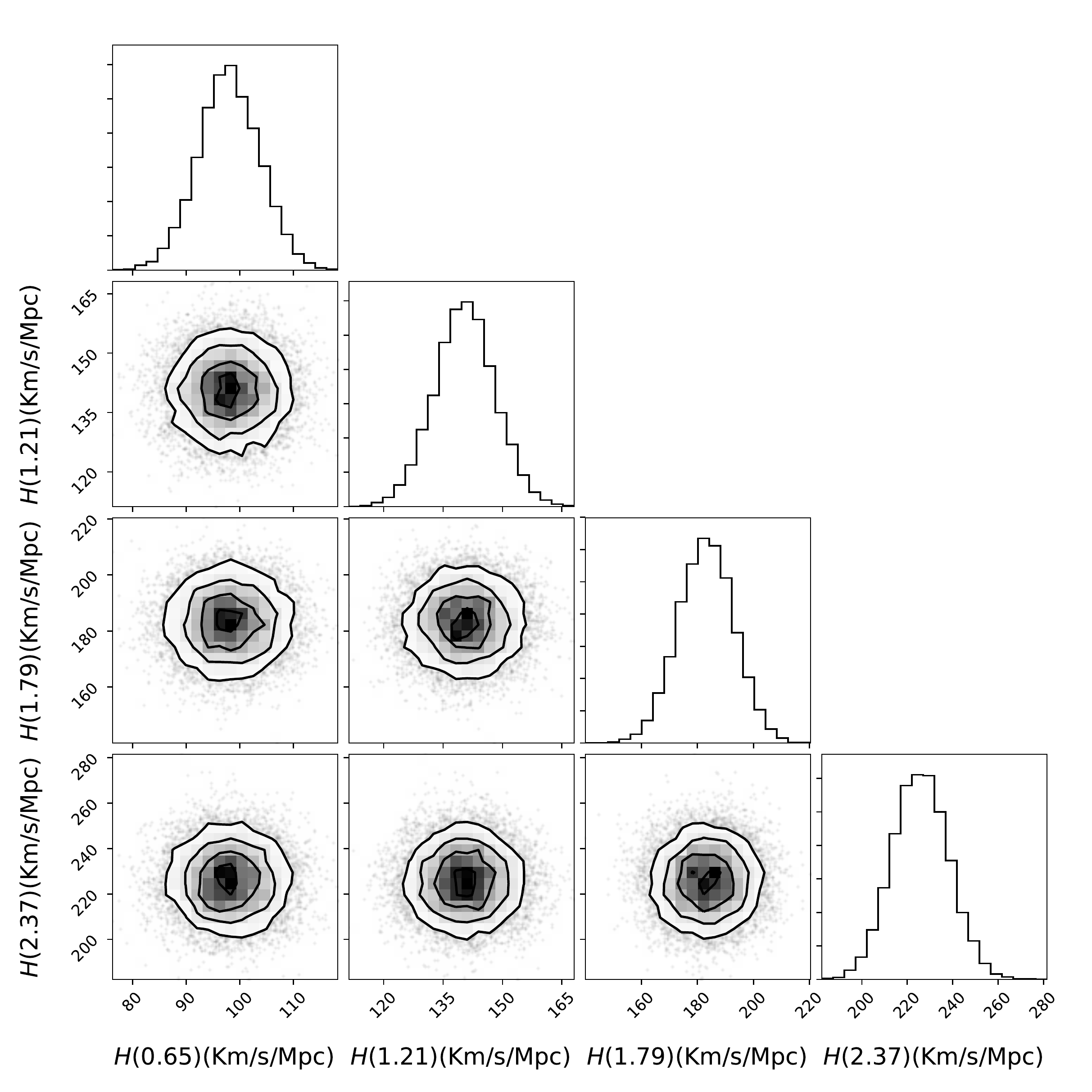}
    \caption{The probability distributions of derived Hubble parameters at
    redshifts $z=0.65$, 1.21, 1.79 and 2.37 from 500 simulated FRBs.
    Contours represent the 1$\sigma$, 2$\sigma$ and 3$\sigma$ confidence levels.} \label{fig2a}
\end{figure}

\begin{figure}
    \centering
    \includegraphics[height=7.5cm,width=10.5cm]{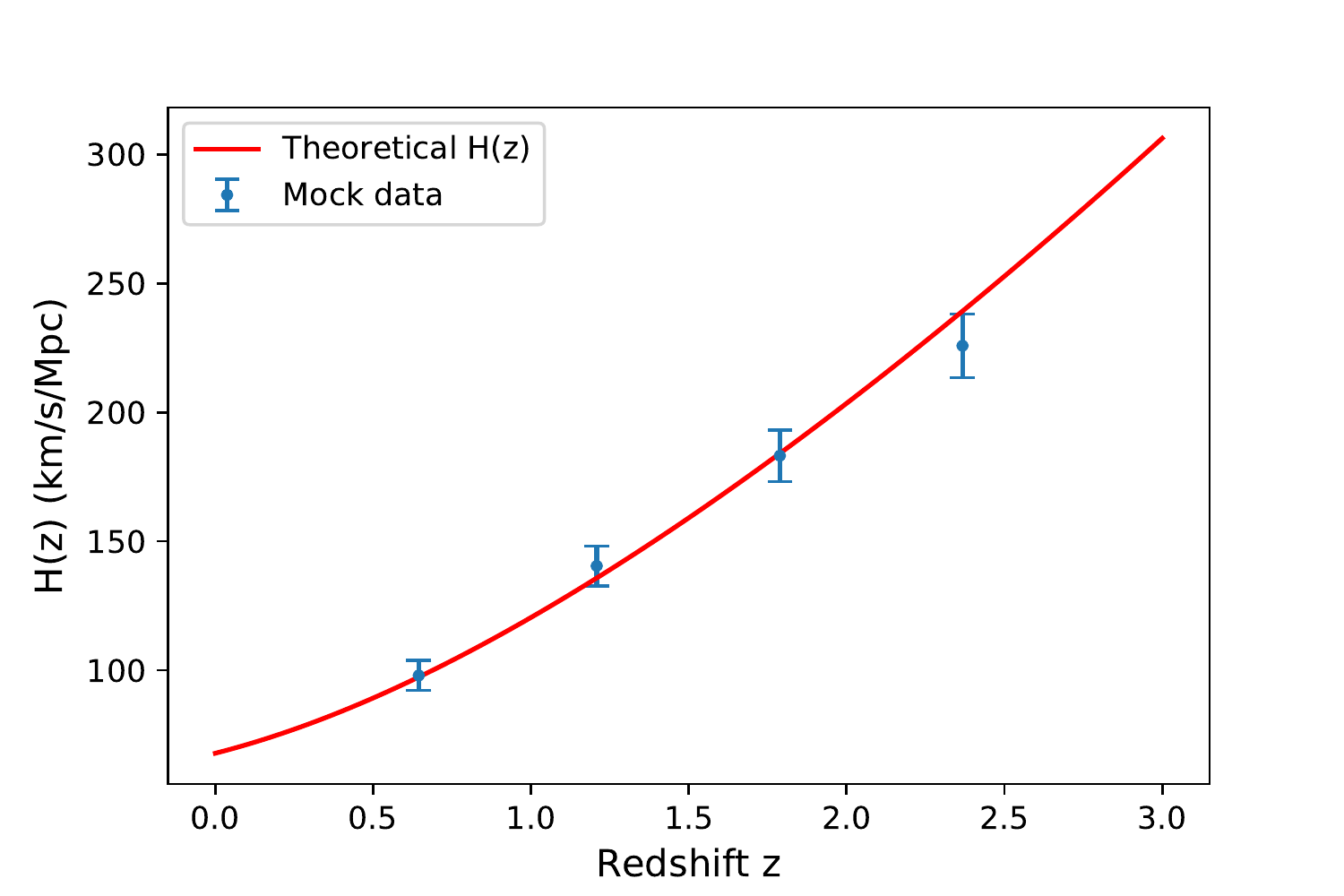}
    \caption{$H(z)$ and 1$\sigma$ errors derived from 500 mock
        FRBs are shown as blue dots. The red line represents the theoretical $H(z)$ function. The deviation between
        the simulated value of $H(z)$ and the theoretical one is about 6\% at $z = 2.4$.
    } \label{fig2}
\end{figure}

Monte Carlo simulation is used to test the efficiency of our method.
Monte Carlo simulation is the easiest way to estimate the
uncertainty of measured $H(z)$ and its dependence on the number of
FRBs. Flat $\Lambda$CDM cosmology with parameters
$\Omega_b=0.0493$, $\Omega_m=0.308$, $\Omega_\Lambda=1-\Omega_m$,
and $H_0=67.8 ~{\rm km/s/Mpc}$ is assumed \citep{Planck2016}. The
redshift distribution of FRBs is assumed as $f(z)\propto ze^{-z}$ in
the redshift range $0<z<3$ \citep{Yu2017}. $H(z)$ is derived as
follows. (i) Simulating a data set $(z,\rm
DM_{E},\sigma_{DM_{total}})$ using equations (\ref{sigma_tot}) and
(\ref{eq_DM_E}). (ii) Then we calculate the Hubble parameter $H(z)$
and $\sigma_{H(z)}$ using above method. (iii) Last, we test our
method using the Kolmogorov-Smirnov (K-S) test. The left panel of
Figure \ref{fig1} shows the 500 mocked $\rm DM_{E}$ data using Monte
Carlo simulation and the theoretical $\rm DM_{E}$ function. The
right panel gives the five binned average $\rm DM_{E}$ with error
about $20 ~{\rm pc/cm^3}$. The average redshift $\langle z \rangle$
and $\langle \rm DM_{E} \rangle$ can be obtained. Then, Hubble
parameter $H(z)$ is derived using equations (\ref{eq_Hz}) and
(\ref{eq_dDM_E2}). In Figure \ref{fig2a}, we give the confidence
regions of $H(z)$ for different redshifts. The best-fit values of
$H(z)$ with 1$\sigma$ errors are $H(0.65)=97.89^{+5.87}_{-5.57}
~{\rm km/s/Mpc}, H(1.21)=140.46^{+7.57}_{-7.61} ~{\rm km/s/Mpc},
H(1.79)=183.14^{+9.79}_{-9.91} ~{\rm km/s/Mpc},
H(2.37)=225.88^{+12.11}_{-12.29} ~{\rm km/s/Mpc}$. Figure \ref{fig2}
presents the derived Hubble parameters $H(z)$ with 1$\sigma$ errors
and the theoretical $H(z)$ function. The derived $H(z)$ is
consistent with the theoretical $H(z)$. The maximum derivation is
about 6\% at redshift $z= 2.4$. The maximum error of $H(z)$ is
$\sigma_{H(z)} \approx 0.06$ using 500 FRBs, which means the
measured Hubble parameter is reliable.

\begin{figure}
    \centering
    \includegraphics[height=7cm,width=10.5cm]{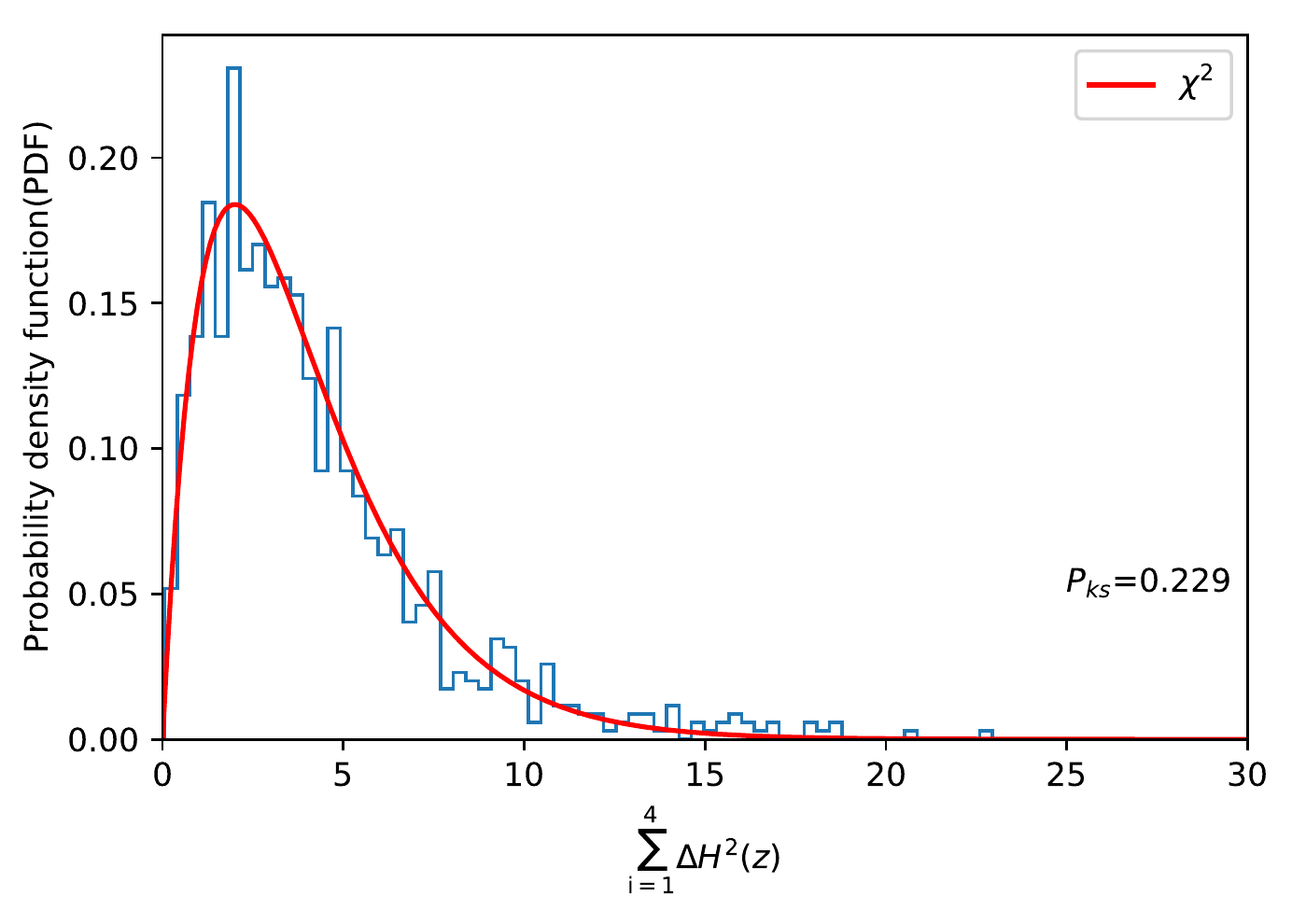}

    \caption{The distribution of the constructed $\mathop{\sum}\limits_{\rm i=1}^{\rm 4}
\Delta H_{\rm i}^2(z)$ from 1000 simulations (blue histogram) and
the probability density function of chi-square ($\chi^2$)
distribution with four degrees of freedom (red line). The $p$-value
of K-S test is 0.229, which supports the two samples are drawn from
the same distribution. } \label{fig3}
\end{figure}

To check internal consistency of the simulation, the
Kolmogorov-Smirnov (K-S) test is used. The $p$-value of K-S test is
considered as the likelihood. In probability theory, if $n$
independent random variables all obey the standard normal
distribution, the sum of squares of the these variables obey the
chi-squared ($\chi^2$) distribution. In order to test the deviation
between the derived $H(z)$ and the theoretical $H_{\rm th}(z)$, we
construct a variable, i.e., $\Delta H(z) = H(z) - H(z)_{\rm th}$,
which obeys normal distribution. Therefore, we compare
$\mathop{\sum}\limits_{\rm i=1}^{\rm 4} \Delta H_{\rm i}^2(z)$ with
the chi-squared distribution. Firstly, we simulate 1000 times and
derive a data set of $H(z)$. Then, the probability density of
$\mathop{\sum}\limits_{\rm i=1}^{\rm 4} \Delta H_{\rm i}^2(z)$ can
be obtained from
\begin{equation}\label{chi}
\Delta \chi^2_c = \boldsymbol{\delta p C^{-1} \delta p} ^{T},
\end{equation}
where $\boldsymbol{C} $ is the covariance matrix of Hubble parameter
$H(z)$, and $\boldsymbol{\delta p}$ is the matrix of difference
between the theoretical and simulated value. Last, we compare it
with the chi-squared distribution, which has the same degree of
freedom. The blue histogram of Figure \ref{fig3} shows the
probability density of $\mathop{\sum}\limits_{\rm i=1}^{\rm 4}
\Delta H_{\rm i}^2(z)$ from 1000 simulations and the red line is the
probability density of chi-squared ($\chi^2$) distribution with four
degrees of freedom. The $p$-value of K-S test is 0.229 which is
larger than 0.05. The two distributions are consistent and the
simulated value and the theoretical value are drawn from the same
distribution.  

Below, the uncertainty of $H(z)$ derived from our method is
discussed. According to equation (\ref{eq_sigma}), the parameter
$\Omega_{b}H_{0}^2 $ will cause uncertainty of derived $H(z)$.
The value of the Hubble constant $H_{0}$ based on Cepheids and
SNe Ia is different as compared to the result based on
the cosmic microwave background (CMB) observations.
\cite{Riess2019} derived the best estimation of $H_{0}=74.03 \pm
1.42 ~\rm km/s/Mpc$ using Cepheid variables, masers in NGC 4258, and
Milky Way parallaxes. In addition, \cite{Freedman2019} found
$H_{0}=69.8 \pm 2.5 ~\rm km/s/Mpc$ based on a calibration of the tip
of the red giant branch applied to SNe Ia. However, $H_{0}=67.8 \pm
0.9 ~\rm km/s/Mpc$ is derived from CMB \citep{Planck2016}. The
difference between them is larger than $4 \sigma$. Here we use the
best constraint on $\Omega_{b}H_{0}^2 $ from Planck CMB data.

Furthermore, when using FRB as cosmological probe, systematic
uncertainty can not be ignored. The dispersion measure induced in
the local vicinity directly by the source may depend the location of
FRB and the properties of its host galaxy. Here we assume that  $\rm
\sigma_{DM_{loc}}=100~ pc/cm^3$ and $\sigma_{\rm DM_{IGM}} = {\rm
DM_{IGM}} / 5\sqrt{z}$ \citep{Kumar2019}. Due to the selection
effect of FRB observations, the DM values observed from different
lines of sight may be different. Therefore, in our simulation, we
average the DM$_{\rm IGM}$ in a small redshift bin.


\section{Conclusions and discussion}\label{sec4}

Fast radio burst is a mysterious astrophysical phenomena that may
provide a new distance probe. In this paper, we propose a new
method to measure Hubble parameter using $\rm DM_{IGM}$ of  FRBs.
Hubble parameter $H(z)$ is a key parameter to reveal the nature of
cosmic expansion and dark energy. Through Monte Carlo simulations,
we used 500 FRBs to measure $H(z)$. K-S test confirms that Monte Carlo simulation is valid
to estimate the uncertainty of measured $H(z)$.
We find that the deviation between the simulated value of $H(z)$ and the theoretical one is
small, and the error is only 0.06. However, the observed DM contains several contributions.
The non-cosmological contributions to DM, and their possible variations
with direction and redshift must be extensively investigated.

Thanks to the high rate of FRBs \citep{Thornton2013}, a large FRB
sample with redshifts can be built in the future. There are
increasing number of observation projects that conduct FRB
observations, such as the Canadian Hydrogen Intensity Mapping
Experiment (CHIME)\citep{The CHIME/FRB Collaboration2019a}, the
Australian Square Kilometre Array Pathfinder
(ASKAP)\citep{Shannon2018}, the Five-hundred-meter Aperture
Spherical Telescope (FAST), which discovered the largest sample of
FRB 121102 \citep{Li2019}. However, the determination of
redshifts for FRBs is still a challenge due to the limitation of
observation technology. In our method to measure $H(z)$ using FRBs,
a large catalog of localized FRBs need to be built up. The Square
Kilometre Array (SKA) can detect FRBs at a rate of $10^3$ sky$^{-1}$
day$^{-1}$ out to redshfit about 3 \citep{Fialkov2017}. If 5\% of
the detected FRBs can be localized, the redshifts of about 10 FRB
host galaxies can be measured per night for a mid-to large-sized
optical telescope \citep{Walters2018}. This indicates that a large
catalog of localized FRBs could be built up.
With a large number of observed FRBs, the reliable measurement of
the Hubble parameter dependence on the redshift will become possible and
will serve as a powerful cosmological probe.

\section*{Acknowledgments}
We thank an anonymous referee for useful suggestions. This work is
supported by the National Natural Science Foundation of China (grant
U1831207).

\end{document}